\documentclass[preprint,onecolumn,showpacs,preprintnumbers,amsmath,amssymb]{revtex4}

\usepackage{mathrsfs}
\usepackage{epsfig}
\usepackage{amsmath}

\newcommand{\bs}{\boldsymbol}
\newcommand{\nn}{\nonumber}
\newcommand{\be}{\begin{equation}}
\newcommand{\ee}{\end{equation}}
\newcommand{\bea}{\begin{eqnarray}}
\newcommand{\eea}{\end{eqnarray}}

\begin{document}

\title{The effect of transverse magnetic correlations on a coupled order parameter: shifted transition temperatures and thermal hysteresis}

\author{K.~L.~Livesey}
\author{R.~L.~Stamps}

\affiliation{School of Physics M013, University of Western Australia, 35 Stirling Hwy, Crawley WA 6009, Australia}

\date{\today}

\begin{abstract}
We use a Green's function method with Random Phase Approximation to show how magnetic correlations may affect electric polarization in multiferroic materials with magnetic-exchange-type magnetoelectric coupling. We use a model spin $\frac{1}{2}$ ferromagnetic ferroelectric system but our results are expected to apply to multiferroic materials with more complex magnetic structures. In particular, we find that transverse magnetic correlations result in a change in the free energy of the ferroelectric solutions leading to the possibility for thermal hysteresis of the electric polarization above the magnetic Curie temperature. Although we are motivated by multiferroic materials, this problem represents a more general calculation of the effect of fluctuations on coupled order parameters.
\end{abstract}


\maketitle

 
\section{Introduction}  
\label{intro}

Early theories that calculated the effect of magnetoelectric (ME) coupling on the spontaneous magnetization $M$ and electric polarization $P$ in so-called multiferroic materials used a Landau free energy formalism \cite{Mitsek,Nedlin,smol}. All terms in the free energy are written as products of the order parameters $M$ and $P$ and must satisfy the symmetry of the paraphase. For example, for a centrosymmetric paraphase, the lowest order magnetoelectric free energy term is $P^2M^2$. Recently, more exotic ME coupling terms have been proposed which also represent free energy invariants, such as a Dzyaloshinskii-Moriya coupling in an antiferromagnet $\bs{P} \cdot \left( \bs{M} \times \bs{L} \right)$ \cite{Fennie,Ederer}, and a spin density wave coupling proposed by Betouras $P \cdot \left(\gamma \nabla \left(\bs{M}^2\right) + \gamma' \left[ \bs{M} \left( \nabla \cdot \bs{M} \right)- \left(\bs{M} \cdot \nabla \right) \bs{M} \right] + \ldots \right)$ \cite{betouras}. These coupling terms explain the multiferroicity of new candidate materials.

However, by writing the ME coupling in terms of the order parameters, the important contribution to the coupling by correlations is ignored. Most obviously, above the lowest ordering temperature (out of the magnetic and the ferroelectric Curie temperatures, $T_{c}^{M}$ and $T_{c}^{P}$ respectively) there can be no ME coupling effects predicted by a Landau theory in zero applied field. In magneto-dielectric materials, magnetic correlations have been shown to have an important effect on dielectric constants \cite{Scott77, Lawes}. In this paper we theoretically study the effect that magnetic correlations may have on spontaneous electric polarization through ME coupling. We find that the electric ordering temperature is shifted, even when it is above the ferromagnetic ordering temperature. We also discover the possibility for thermal hysteresis in $P$ for multiferroic systems when a particular ME coupling term is allowed. While calculations published in the last few years go beyond Landau theory \cite{zhong2,Apost,Wessel,Wessel2}, they fail to identify this effect.

The more general problem of understanding how correlations affect second order phase transitions in systems with coupled order parameters has been approached in the past \cite{Larkin69}, particularly using techniques from renormalization group theory \cite{Aharony73}. For a one dimensional system, the coupled parameters can be calculated exactly \cite{Imry74} but for more general systems this is not the case. Our Green's function technique, which in this paper only includes transverse magnetic correlations, represents a unique and new approach. It also gives meaningful results for all temperatures, unlike the renormalization group method which can only give information on the critical behavior. 

In Sec.~\ref{GFsection} we detail the Green's function technique with Random Phase Approximation and derive a free energy for a ferromagnetic ferroelectric bulk material, which can be utilized to solve for the coupled order parameters $M$ and $P$, as well as other thermodynamic quantities. In Sec.~\ref{results} we provide some results for ME coupling which is both linear and quadratic in the order parameter $P$. We show how the coupling linear in $P$ leads to the possibility for thermal hysteresis by altering the free energy of different local energy minima. In Sec.~\ref{conclude} we summarize the results, discuss the limitations of the current theory and provide an outlook on future work.

\section{Green's function method}
\label{GFsection}

A model ferromagnetic ferroelectric system with so-called ``isotropic" or ``exchange" ME coupling \cite{smol} is treated. The methods presented can be extended to treat multi-sublattice magnets but become much more complicated. We assume that $T_{c}^{M} <T_{c}^{P}$ since this is the case for the majority of multiferroic materials \cite{smol} and also since then the effects of magnetic correlations on $P$ will be more significant. 

The ferroelectric system is modeled using Landau theory for second-order phase transitions with free energy density given by:
\be
F_{FE} = \frac{1}{2} \alpha P^2 + \frac{1}{4} \beta P^4 - E P,
\label{FEfreeE}
\ee
where $E$ is an applied electric field parallel to the spontaneous polarization, $\alpha = A k_{B} (T-T_{c})$, $A$ and $\beta$ are phenomenological constants, $k_{B}$ is Boltzmann's constant and $T$ is temperature. $T_{c}$ is the ferroelectric Curie temperature in the absence of ME coupling. Landau theory is valid near phase transitions so we assume that $T_{c}^{M}$ and $T_{c}^{P}$ are sufficiently close for the results to be valid. $P$ is treated as a scalar quantity. Its direction relative to the magnetization is not relevant in this simple model.

We aim to write a free energy for the magnetic system with ME coupling to combine with Eq.~\eqref{FEfreeE} in order to solve for both order parameters simultaneously. We start with a microscopic Hamiltonian and use a Green's function technique (GFT) with a Random Phase Approximation (RPA) to derive the free energy.

The starting spin Hamiltonian is
\bea
\hat{H} &=& - h \sum_{i} \hat{S}_{i}^{z}  -\frac{1}{2} \left( J+ \Gamma P+\gamma P^2 \right) \sum_{\langle i,j \rangle}  \hat{\bs{S}}_{i}^{} \cdot \hat{\bs{S}}_{j}^{}  \nn \\
&\sim& - h \sum_{i} \hat{S}_{i}^{z}  - \frac{1}{2} \left( J+ \Gamma P+\gamma P^2 \right) \sum_{\langle i,j \rangle} \left( \hat{S}_{i}^{+} \hat{S}_{j}^{-} \right. \nn \\
&& \left. +  \hat{S}_{i}^{z} \langle S_{}^{z} \rangle  + \hat{S}_{j}^{z} \langle S_{}^{z} \rangle + \langle S_{}^{z} \rangle^2 \right) ,
\label{SpinHam2}
\eea
where the first term is the Zeeman interaction with $h=g \mu_{B} H_{0}$ and $H_{0}$ is an applied magnetic field. The second term represents the exchange interaction with a Taylor series expansion for the weak contribution from electric polarization $P$. Sushkov \emph{et al.} recently used a Hamiltonian of similar form to describe how the magnetic exchange interactions together with magnetostriction in $R$Mn$_2$O$_5$ ($R$=Y,Bi) couple strongly the magnetic system to a soft phonon mode associated with a spontaneous electric polarization \cite{sushkov08}. The sum is over nearest neighbor pairs of spins at sites $i$ and $j$ and $\hat{S}_{i}^{\pm} = \hat{S}_{i}^{x} \pm i \hat{S}_{i}^{y}$. The constants $\Gamma$ and $\gamma$ describe the ME coupling strength that is linear and quadratic respectively in the spatially averaged order parameter $P$. We assume for a ferromagnet that all sites $i$ are equivalent.

Longitudinal spin terms $S_{i}^{z} S_{j}^{z} $ are ignored in Eq.~\eqref{SpinHam2}, but transverse terms $ S_{i}^{+}S_{j}^{-}$ are kept. The reason for ignoring the longitudinal correlations is that RPA is known to produce spurious solutions for $\langle S_{i}^{z} S_{j}^{z} \rangle$ \cite{tahirbook, tahir67}. Although this does not cause significant problems when calculating the magnetization $M \equiv \langle S_{}^{z} \rangle$, it does introduce significant error in the subsequent calculation of the free energy. More advanced decoupling procedures may be used to gain better results for the longitudinal correlations \cite{tahirbook} but here we demonsrate possible effects of including transverse correlations as a first step. It should be noted that Callen decoupling \cite{Callen} has the same problems as RPA with regard to longitudinal correlations and was specifically designed to improve RPA only when treating weak single-site anisotropies. We ignore anisotropies here since we will do the calculation for a spin $\frac{1}{2}$ system.

We define retarded Green's functions \cite{bogolyubovGF,Zubarev}
\be
G_{ij} (t) \equiv \langle \langle S_{i}^{+} ; S_{j}^{-} \rangle \rangle \equiv -i \theta (t) \langle [ S_{i}^{+} (t), S_{j}^{-} ] \rangle ,
\label{defnRetarded}
\ee
where the square brackets indicate a commutator such that $[A,B] = AB-BA$, the single angled brackets indicate a statistical thermal average and the $\theta(t)$ function is a unit step function. The time Fourier transform of Eq.~\eqref{defnRetarded} is given by
\be
G_{ij} (\omega) \equiv \langle \langle  S_{i}^{+}; S_{j}^{-} \rangle \rangle_{\omega} = \frac{1}{2 \pi} \int_{-\infty}^{+\infty}  d t \phantom{1} G_{ij} (t) e^{-i \omega t} .
\label{TimeFT}
\ee
Then the equation of motion for the time Fourier-transformed Green's function is
\be
\omega G_{ij} (\omega) = \frac{ \delta_{ij} }{2 \pi} \langle [S_{}^{+},S_{}^{-}] \rangle + \langle \langle [S_{i}^{+}, H]; S_{j}^{-} \rangle \rangle_{\omega} ,
\label{EqMotionFreq}
\ee
where $\delta_{ij}$ is the discrete Kronecker delta function. Substituting Eq.~\eqref{SpinHam2} into Eq.~\eqref{EqMotionFreq} gives
\bea
\omega G_{ij}(\omega) &=& \frac{\delta_{ij}}{2 \pi} \langle2 S_{}^{z} \rangle  
\label{GFeqmotion} \\
&&+ \left( h + z (J+\Gamma P + \gamma P^2) \langle S_{i}^{z} \rangle \right) G_{ij}(\omega) 
\nn \\
&&- (J+\Gamma P + \gamma P^2) \sum_{l} \langle \langle  S_{i}^{z} S_{l}^{+} ; S_{j}^{-}  \rangle \rangle_{\omega}^{}   \nn ,
\eea
where $z$ is the number of nearest neighbors to a site. The sum over $l$ is over the nearest neighbors to site $i$. 

The last term in Eq.~\eqref{GFeqmotion} represents a higher order Green's function which can be approximated using RPA:
\be
\langle \langle  S_{i}^{z} S_{l}^{+} ; S_{j}^{-}  \rangle \rangle_{\omega}^{}    \sim   \langle S^{z} \rangle  \langle \langle S_{l}^{+}; S_{j}^{-} \rangle \rangle_{\omega}^{} = \langle S^{z} \rangle G_{lj} (\omega)  
\label{RPA}
\ee
in order to obtain a solution for $G_{ij} (\omega)$. We perform a spatial Fourier transform
\bea
G(\omega,\bs{k}) &=& \sum_{\bs{r}_{i}-\bs{r}_{j}} G_{ij} (\omega) e^{-i \bs{k} \cdot (\bs{r}_{i}-\bs{r}_{j})} \\
\label{GFFT}
G_{ij}(\omega) &=& \frac{1}{N} \sum_{\bs{k}} G(\omega, \bs{k} ) e^{i \bs{k} \cdot (\bs{r}_{i} - \bs{r}_{j}) } ,
\eea
where the sum over $\bs{r}_{i}-\bs{r}_{j}$ is over all displacements from site $i$ to site $j$, and solve Eq.~\eqref{GFeqmotion}:
\bea
G(\omega, \bs{k}) &=& \frac{\langle S^{z} \rangle}{\pi (\omega - \omega_{\bs{k}})} , 
\label{GFsoln} \\
\omega_{\bs{k}} &=& (J+ \Gamma P + \gamma P^2) z \langle S^{z}\rangle (1-\gamma_{\bs{k}}) +h .
\label{GFdisp}
\eea
The structure factor $\gamma_{\bs{k}}  \equiv \frac{1}{z} \sum_{i} e^{i \bs{k} \cdot \bs{a}_{i} } $, where $\bs{a}_{i}$ is the displacement from the reference site to its neighbor $i$, should not be confused with the coefficient for ME coupling $\gamma$. 

Eq.~\eqref{GFdisp} gives the dispersion relation for magnons within the RPA. The average electric polarization can be seen to alter the dispersion and application of an electric field will be able to shift the frequencies via the isotropic ME coupling. The possibility to tune spin wave dispersion using electric fields may have potential application in spin wave logic devices \cite{SWlogic,deSousaSW}. If our model allowed for fluctuations in the electric polarization, rather than just the magnetization, then the resonant modes may be ``electromagnons" with a dual magnetic/electric nature.

The transverse correlation function between neighboring spins $i$ and $j$ is calculated from Eq.~\eqref{GFsoln} using the Spectral Theorem \cite{Zubarev} with the aid of complex variable methods:
\bea
\langle S_{j}^{-} S_{i}^{+} \rangle &=& i \lim_{ \substack{\epsilon \to 0 \\ t \to 0} } \frac{1}{N} \sum_{\bs{k}} e^{-i \bs{k} \cdot (\bs{r}_{i}-\bs{r}_{j}) }  \int_{-\infty}^{\infty} d \omega \frac{e^{-i \omega t}}{e^{\frac{\omega}{k_{B} T}}-1}  \nn \\
&& \times \left[  \frac{\langle S^{z} \rangle}{\pi (\omega+i \epsilon-\omega_{\bs{k}})} - \frac{\langle S^{z} \rangle}{\pi (\omega-i \epsilon-\omega_{\bs{k}})}  \right] \nn \\
&=& \frac{1}{N} \sum_{\bs{k}} \frac{ 2 \langle S_{}^{z} \rangle e^{-i \bs{k} \cdot (\bs{r}_{i}-\bs{r}_{j}) } }{ \left( e^{\frac{\omega_{\bs{k}}}{ k_{B} T }}-1 \right) } .
\label{TransCorr}
\eea

For spin~$\frac{1}{2}$ the magnetization is given by \cite{TahirKheli,bogolyubovGF} 
\be
M \equiv \langle S_{}^{z} \rangle =  \frac{1}{2} \left( 1 +2 \Phi \right)^{-1} , 
\label{Meqn}
\ee
where $ \Phi= \frac{1}{N}  \sum_{\bs{k}} ( e^{\frac{\omega_{\bs{k}} }{k_{B} T }} -1 )^{-1} $. For general spin the result is \cite{Callen}
\be
M = \frac{ (S-\Phi) (1+\Phi)^{2S+1} + (S+1+\Phi) \Phi^{2S+1} }{ (1+\Phi)^{2S+1} - \Phi^{2S+1} } .
\label{CallenEq}
\ee

Having found a self-consistent equation for $M$, we need also to find an equation for $P$ using the free energy. We follow the workings in Appendix A of Ref.~\cite{Froebrich} to derive the free energy.

The expectation value of the single-site Hamiltonian, derived from Eqs.~\eqref{SpinHam2} and \eqref{TransCorr}, gives the intrinsic energy $\mathscr{E}$ per magnetic lattice site:
\bea
\mathscr{E}= \langle H_{i} \rangle &=&  -h \langle S^{z} \rangle - \frac{z}{2}  (J+ \Gamma P + \gamma P^2) \langle S^{z} \rangle 
\nn \\ 
&& \times \left[ \langle S^{z} \rangle +  \frac{1}{N} \sum_{\bs{k}} \frac{\gamma_{\bs{k}}}{(e^{\frac{\omega_{\bs{k}}}{k_{B} T}}-1)} \right] .
\label{internalE}
\eea
From the intrinsic energy, we can derive an expression for the free energy $F$ by making use of the relations:
\bea
F &=& \mathscr{E} - T S  ,\\
S &=& \left( \frac{ \partial F }{ \partial T } \right)_{M} , 
\label{entropy}
\eea
where $S$ is the entropy. Rearranging these we obtain
\be
F(T) = \mathscr{E}(0) - T \int_{0}^{T} d \tau \frac{ \mathscr{E}(\tau)-\mathscr{E}(0) }{ \tau^2 } .
\label{FfromE}
\ee
This free energy is not a function of $M\equiv \langle S^{z} \rangle$ since $M$ must be constant in the definition of $S$ which we use [Eq.~\eqref{entropy}]. Substituting Eq.~\eqref{internalE} into Eq.~\eqref{FfromE} we obtain the free energy. Adding it to the ferroelectric free energy [Eq.~\eqref{FEfreeE}], the total free energy per magnetic unit cell of volume $V$ is given by 
\bea
F_{\textrm{GFT}} &=& V \left( \frac{1}{2} \alpha P^2 + \frac{1}{4} \beta P^4 - E P \right) 
\label{freeEmagGF} \\
&& -h \langle S^{z} \rangle - \frac{z}{2} (J+\Gamma P+\gamma P^2) \langle S^{z} \rangle  
\nn \\ 
&&  \times \left[ \langle S_{}^{z} \rangle +  \frac{ 2 k_{B} T  }{ N } \sum_{ \bs{k} } \frac{  \gamma_{\bs{k}}  }{ \omega_{\bs{k}} } \ln \left(1 - e^{- \frac{ \omega_{\bs{k}} }{ k_{B} T }} \right) \right] . \nn
\eea
This expression is true for general spin. The last term in Eq.~\eqref{freeEmagGF} represents the contribution from the transverse correlations and at low temperatures gives the free energy of a gas of noninteracting magnons, proportional to $k_{B} T \sum_{\bs{k}} \ln (1 - e^{- \frac{ \omega_{\bs{k}} }{ k_{B} T }} )$. A solution for $P$ can be found by numerically minimizing Eq.~\eqref{freeEmagGF}. 

To find $M(T)$ and $P(T)$ we use an iterative procedure, starting at low temperatures and using $M_{0} = \pm \frac{1}{2}$ as an initial point. Substituting $M_{0}$ into Eq.~\eqref{freeEmagGF} and minimizing, we obtain $P_{1}$. Substituting $P_{1}$ into the right hand side of Eq.~\eqref{Meqn}, we obtain $M_{1}$. This process is repeated $n$ times until there is no longer a change in $M_{n}$ and $P_{n}$ to the required precision. For higher temperatures, the most useful starting point for iteration is the solution already found for lower temperatures.

We will compare our results to those found using mean field theory (MFT) in order to examine the effect that the transverse magnetic correlations have on the solution. MFT corresponds to ignoring the transverse correlation term between neighbors, $\hat{S}_{i}^{+} \hat{S}_{j}^{-}$, in Eq.~\eqref{SpinHam2} and thus corresponds to reducing the problem to a single particle problem. The resulting free energy for spin~$\frac{1}{2}$ can be found simply using the magnetic partition function $Z$:
\bea
F_{MFT} &=& F_{FE} - k_{B}T \ln Z  \nn \\
&=& F_{FE} -k_{B}T \ln \left( \sum_{S_{i}^{z} = \pm \frac{1}{2}} \langle S_{i}^{z} | e^{-\frac{ \hat{H}_{i}}{k_{B}T}} | S_{i}^{z} \rangle \right) \nn \\
&=& V \left( \frac{1}{2} \alpha P^2 + \frac{1}{4} \beta P^4 - E P \right) 
\label{freeE_MFT}  \\
&& - \frac{z}{2} (J+\Gamma P+\gamma P^2) \langle S^{z} \rangle^2  
\nn \\ 
&& -k_{B} T \ln \left( \cosh \left( \frac{h + z \langle S_{}^{z} \rangle (J+\Gamma P+\gamma P^2)}{2 k_{B} T} \right) \right) . \nn
\eea
This free energy may be minimized with respect to both $M$ and $P$, unlike Eq.~\eqref{freeEmagGF}, to solve for the order parameters at a given temperature $T$.

\section{Results}
\label{results}

\subsection{Changes to critical temperatures}

\begin{figure}[htbp]
\centering
\includegraphics[width=7cm]{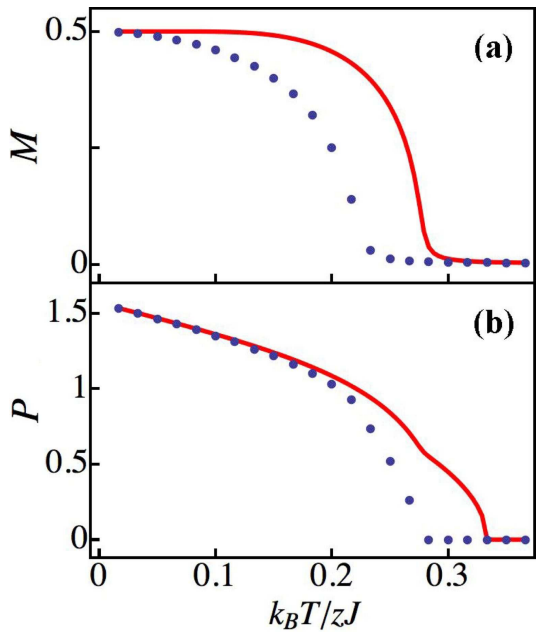}	
\caption[The magnetisation (a) and electric polarisation (b) of a ferromagnetic ferroelectric with isotropic ME coupling that is quadratic in $P$, calculated using mean field theory (solid lines) and using our Green's function method (dots).]{\label{IsoQuadMEcoupling}(Color online) The magnetization (a) and electric polarization (b) plotted as a function of normalised temperature $k_{B} T/z J$ of a ferromagnetic ferroelectric with isotropic ME coupling that is quadratic in $P$, calculated using mean field theory (solid lines) and using Green's function method (dots). The parameters used are $\alpha = A k_{B} (T - T_{c}^{})$, $k_{B} T_{c}/z J = 1/3$, $A=\beta=1$, $\Gamma=0$, $\gamma = 0.05 zJ$ and $h=0.0017 z J$. A simple cubic lattice is assumed.}
\end{figure}

In Fig.~\ref{IsoQuadMEcoupling} the magnetization (panel a) and electric polarization (panel b) are plotted as a function of normalized temperature $k_{B} T/z J$ for a $S=\frac{1}{2}$ ferromagnetic ferroelectric with isotropic magnetoelectric coupling that is quadratic in $P$. The MFT results (Eq.~\eqref{freeE_MFT}) are shown by the solid lines and the GFT results are shown by the dots. The material parameters used are $\alpha = A k_{B} (T - T_{c}^{})$, $k_{B} T_{c}/z J = 1/3$, $A=\beta=1$, $\Gamma=0$ and $\gamma/zJ = 0.05$. The coupling strength $\gamma$ is unphysically large but allows the effects of the isotropic coupling to be seen clearly. Also, a small applied field given by $h/zJ = 0.0017$ is applied in order that the magnetic Green's function is defined at all temperatures.
 
 From Fig.~\ref{IsoQuadMEcoupling}(a), the GFT gives a value for the Curie temperature $T_{c}^{M}$ lower than the mean field value. The reason is that an introduction of correlations results in less thermal energy being necessary to flip spins, thereby lowering the critical temperature. From Fig.~\ref{IsoQuadMEcoupling}(b), we see that the Green's function with RPA treatment for the spin system gives a critical ferroelectric temperature which is also lower than the mean field prediction. Most stikingly, the magnetic transverse correlations affect the ferroelectric system above the magnetic Curie temperature and cause a reduction in the spontaneous polarization.
 
 In Fig.~\ref{IsoQuadMEcoupling} only the solutions corresponding to $M>0$ and $P>0$ are shown. However, there are four solutions which correspond to local energy minima for the ferromagnetic ferroelectric system in the four different quadrants of the phase space given by $\{ M,P \}$. Because of the symmetry breaking applied field $h>0$, the two solutions with $\{M>0, P>0\}$ and $\{M>0,P<0\}$ (call them $S_{++}$ and $S_{+-}$ respectively) are degenerate and correspond to the lowest energy solutions for isotropic coupling that is quadratic in $P$. Applying a symmetry-breaking positive electric field breaks the degeneracy and makes $S_{++}$ the equilibrium solution for all temperatures. This is not the case when we consider a multiferroic system with isotropic magnetoelectric coupling that is linear in $P$.

\subsection{Thermal hysteresis}

\begin{figure}[h]
\centering
\includegraphics[width=12cm]{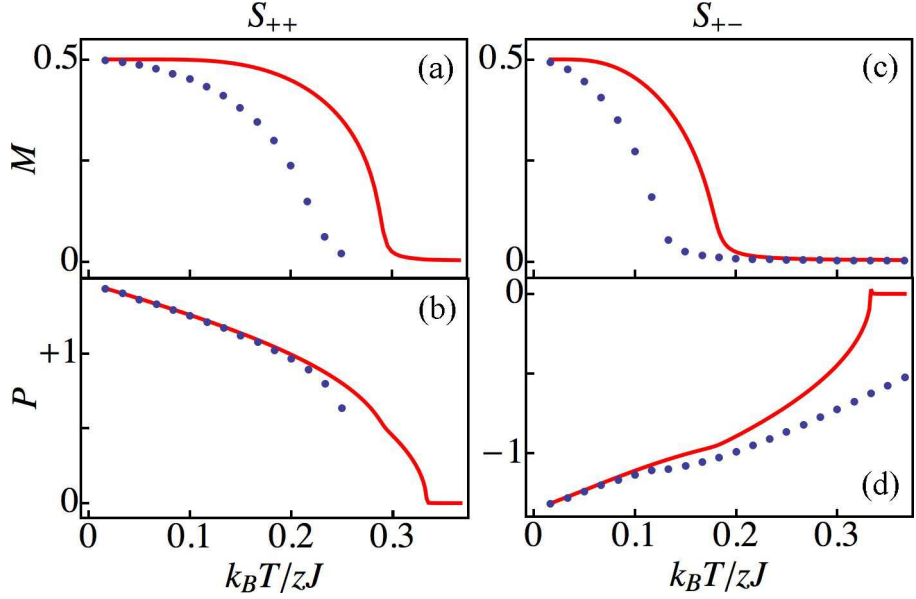}	
\caption{\label{2solns}(Color online) Two solutions for $\{M,P\}$ in a ferromagnetic ferroelectric with isotropic ME coupling that is linear in $P$, calculated using mean field theory (solid lines) and Green's function methods (dots). Panels (a) and (b) show the magnetization and polarization as a function of temperature corresponding to solution $S_{++}$, which has positive $P$. Panels (c) and (d) show the magnetization and polarization as a function of temperature corresponding to solution $S_{+-}$, which has negative $P$. The material parameters used are $\alpha = A k_{B} (T - T_{c}^{})$, $k_{B} T_{c}/z J = 1/3$, $A=\beta=1$, $\Gamma=0.05 z J$, $\gamma = 0$ and $h=0.0017 z J$. A simple cubic lattice is assumed.}
\end{figure}

In Fig.~\ref{2solns} we show two solutions $S_{++}$ (both $M$ and $P$ positive) and $S_{+-}$ ($M$ positive and $P$ negative) as a function of normalized temperature $k_{B} T/ zJ$ when the ME coupling is linear in $P$. $S_{++}$ has magnetization plotted in panel (a) and polarization in panel (b). $S_{+-}$ has magnetization plotted in panel (c) and polarization in panel (d). The solid lines show the MFT results and the dots show the GFT results. The parameters used are given in the figure caption. The coupling linear in $P$ is extremely rare in ferromagnets since it relies on broken inversion symmetry. However, it may exist in frustrated spin structures, as are typical for multiferroic materials. 

 The most interesting deviation from the MFT results is that $S_{++}$ no longer exists above a temperature of $k_{B}T/ z J \sim 0.25$ when transverse correlations are included (see Fig.~\ref{2solns}(a) and (b)). This local free energy minima ceases to exist and there is a discontinuous ferroelectric transition from one solution to the other. To explain this result, we need to consider the free energy of the solutions. 
 
\begin{figure}[h]
\centering
\includegraphics[width=7cm]{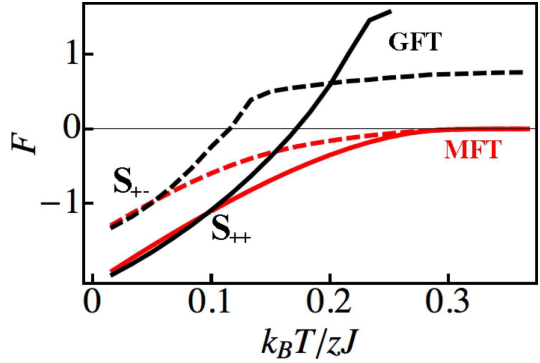}	
\caption{\label{freeEplots}(Color online) The free energy corresponding to solutions $S_{+-}$ (dashed lines) and $S_{++}$ (solid lines) for the ferromagnetic ferroelectric with isotropic ME coupling. The GFT predictions are shown in black and the MFT predictions are shown in red. Material parameters used are the same as those in Fig.~\ref{2solns}.}
\end{figure}

 In Fig.~\ref{freeEplots} we show the free energy of the two solutions illustrated in Fig.~\ref{2solns} as a function of normalized temperature. $S_{++}$ is shown by solid lines and $S_{+-}$ is shown by dashed lines. The MFT free energies [Eq.~\eqref{freeE_MFT}] of the two solutions are shown in red and converge at high temperatures. The GFT results [Eq.~\eqref{freeEmagGF}] are shown in black. When correlations are ignored, the MFT results show that $S_{++}$ is always the lowest energy solution. This is because the coupling gives rise to an effective symmetry-breaking electric field $E_{\textrm{eff}} = \frac{\Gamma z M^2}{2 V} >0$ (see Eq.~\eqref{freeE_MFT}). This is also the case in the GFT for low temperatures. However, at higher temperatures where the transverse correlations become large, the situation is different.
 
 Solutions $S_{++}$ and $S_{+-}$ have different effective exchange interactions $J_{\textrm{eff}} = J+ \Gamma P$. In consequence, the contributions to the free energy due to transverse correlations are different in each case. The energy associated with the correlations can be deduced by subtracting the MFT free energy from that found using the GFT (see Fig.~\ref{freeEplots}). Solution $S_{+-}$ has a lower exchange energy associated with correlations and also a lower Curie temperature than $S_{++}$. This means that at $k_{B}T/zJ \sim 0.2$ the lowest free energy state changes from $S_{++}$ to $S_{+-}$ (see Fig.~\ref{freeEplots}). Also, for $k_{B} T/zJ >0.25$, the transverse correlations add an additional energy that makes $S_{++}$ unstable.
 
 An intriguing consequence is the possibility of thermal hysteresis. The state of the system depends not only on its temperature but also on the temperature it has had in the past. If the system is heated to above $k_{B} T/z J \sim 0.25$ then it must have solution $S_{+-}$ since $S_{++}$ vanishes. If it is then cooled back down to below $k_{B} T/zJ \sim 0.2$, the system may remain in solution $S_{+-}$ even though it is now a metastable solution (see Fig.~\ref{freeEplots}). There is a probability that thermal fluctuations may take the system back to state $S_{++}$ with a characteristic time for switching $\tau$ which depends on the temperature and the height of the energy barrier separating the solutions. If $\tau$ is quite large then we have a long-lived two state system which may be implemented for information storage, provided the thermal hysteresis occurs in a region spanning room temperature. The application is to thermally assisted ferroelectric recording, where the coupling to magnetic correlations has enabled there to be thermal hysteresis of the electric polarisation $P$. As far as multiferroics are concerned, the only experimental evidence of thermal hysteresis in $P$ that we are aware of as yet is at commensurate-incommensurate spin density wave transitions in DyMn$_2$O$_5$ at low temperature \cite{delaCruz}. 

We must question how our results would be changed if longitudinal correlations were included in the theory. We expect our result of thermal hysteresis to still hold because of the following argument. At very low temperatures, $S_{++}$ is the stable solution of the system and both the MFT and GFT agree on this result (see Fig.~\ref{freeEplots}). At high temperatures $T> T_{++}^{M}, T_{+-}^{M}$ and with $h \to 0$, $\langle S_{i}^{z} S_{j}^{z} \rangle = \langle S_{i}^{x} S_{j}^{x} \rangle$ due to symmetry considerations. So we can expect that the difference in the free energy of the two solutions would have the same sign and would be larger if longitudinal correlations were included. Therefore, at high temperatures the result that $S_{+-}$ is the stable solution seems robust. With the low temperature and high temperature results assumed correct, the system necessarily switches from one solution to another at intermediate temperatures.

\section{Conclusion}
\label{conclude}

We have presented a model which goes beyond Landau theory of second order phase transitions to examine the thermodynamic properties of a system with coupled order parameters. We use a Green's function theory with Random Phase Approximation to include the effect of transverse magnetic correlations on the free energy of a multiferroic system with magnetization $M$ and polarization $P$. We find that the ferroelectric transition temperature is shifted as compared with mean field predictions that ignore magnetic correlations. This has implications when estimating ME coupling strength in experiments that measure a shift in ferroelectric polarization on application of a magnetic field at finite temperature. We also show that if a ME coupling that is linear in $P$ is symmetry-allowed in a system, then there is the possibility for thermal hysteresis of $P$. Which free energy minimum represents the lowest energy state switches from one solution to another at finite temperature due to each solution having different exchange energy contributions.

Our method has the advantage of being well-defined at all temperatures, unlike renormalization group methods. However, longitudinal magnetic correlations are ignored to simplify the theory and also any fluctuations in the ferroelectric system are ignored as a first approximation. More sophisticated decoupling procedures exist to include longitudinal magnetic correlations without creating spurious results for the free energy \cite{tahirbook}. Also, it is possible to go beyond the Landau treatment of the ferroelectric system using, say, a Hubbard model which is suitable for displacive ferroelectrics (see for example Ref.~\cite{ishihara94}). 

If the method were to be used to model a specific multiferroic material, then an extension to treat multi-sublattice magnets would be necessary. This would be possible using existing Green's function methods for antiferromagnets (see for example Ref.~\cite{Anderson} or \cite{jensen06afm}).

\begin{acknowledgments}
We acknowledge funding from the Hackett Student Fund, UWA Completion Scholarship, Seagate Technologies and the Australian Research Council. KLL thanks Prof. Peter Fr\"obrich for useful advice.
\end{acknowledgments}


 \end{document}